\documentclass[superscriptaddress,aps,preprint,amsmath,amssymb,floatfix]{revtex4-1}
\usepackage{graphicx}
\usepackage{color}
\usepackage{bm}
\usepackage{hyperref}
\usepackage{ccaption}%

\newcommand{\bk}[0]{{\bf k} } 
\newcommand{\km}[0]{{\bf k} } 
\newcommand{\bp}[0]{{\bf p} }
\newcommand{\bq}[0]{{\bf q} }

\newcommand{\bQ}[0]{{\bf Q} }

\usepackage{ulem}

\usepackage{caption,newfloat}
\DeclareCaptionType{InfoBox}

\begin{document}

\begin{center}

{\large{\bf
{Quantum critical metals and loss of quasiparticles}
}
}
\\[0.5cm]


Haoyu Hu$^{1,2,\ast}$,
Lei Chen$^{1,\ast}$,
Qimiao Si$^{1,\ast}$

{\em $^1$Department of Physics and Astronomy, 
Extreme Quantum Materials Alliance,
Smalley-Curl Institute,
Rice University,
Houston, TX 77005, USA}
\\

{\em $^2$Donostia International Physics Center, P. Manuel de Lardizabal 4, 20018 Donostia-San Sebastian, Spain}
\\


\end{center}

\vspace{0.5cm}
{\bf
Strange metals develop near quantum critical points  in a variety of strongly correlated systems. Some of the issues that are central to the field
include how the quantum-critical state loses quasiparticles, how it drives superconductivity, and to what extent the strange-metal physics in different classes of correlated systems are interconnected. In this Review, we survey some of these issues from the vantage point of heavy fermion metals. We will describe the notion of Kondo destruction and how it leads to several crucial effects. 
These include a transformation of the Fermi surface from large to small 
when the system is tuned across the quantum-critical point, a loss of quasiparticles everywhere on the Fermi surface 
when it is perched at the quantum-critical point, and a dynamical Planckian scaling in various physical properties including charge responses. 
We close with a discussion about the connections between the strange-metal 
physics in heavy fermion metals and its counterparts in the cuprates and 
other correlated materials.
}
\vspace{0.6cm}

\noindent E-mail: \vspace{0.6cm}
huhaoyu314@gmail.com; lc73@rice.edu; qmsi@rice.edu

\newpage





\section{Introduction}

Large classes of quantum materials host
strongly correlated 
electrons \cite{Kei17.1,Pas21.1} and many of them feature 
unconventional superconductivity. One connection among the strongly correlated systems is illustrated in Fig.\,\ref{fig:Over}(a){, which shows the superconducting transition temperature $T_c$ and the effective Fermi temperature $T_0$, 
{the temperature for Fermi degeneracy},
for various strongly-correlated superconductors}. 
The ratio of $T_c/T_0$ is several percent,
with each 
temperature scale spanning about three decades. 
This qualifies these systems as 
high-$T_c$ superconductors,
given that this ratio is about two orders of magnitude smaller in conventional 
superconductors. 
Another connection lies in their normal states 
at temperatures above the superconducting transition temperature (so, when $T>T_c$), which are often strange metals 
that have an electrical resistivity that is linear in temperature, and
a slew of 
other 
exotic properties. 

The link between the strange-metal normal state and unconventional superconductivity in heavy fermion systems{, which are characterized by 
electronic excitations whose effective masses 
are orders of magnitude larger than the free electron mass}, is particularly striking. Indeed, heavy fermion metals represent a prototype setting
in which quantum critical metallicity has been elucidated \cite{KirchnerRMP}, in part because $T_c$ is relatively small in absolute magnitude 
in these materials, so it opens up a large window of temperature over which the strange-metal properties can be explored. 
These systems 
often 
possess antiferromagnetic (AF) correlations.
The existence of heavy fermion superconductors is a venerable topic, and
this material family has now grown to about $50$ members. In contrast,
the  strange-metal behavior and its association with quantum criticality 
have only been the focus relatively recently. 

It is natural for quantum criticality to drive unusual properties \cite{Mathur98}.
Indeed, as a system is tuned towards 
its quantum-critical regime at a given low (but nonzero) temperature, the 
entropy is expected to be maximized \cite{ZhuGarst.03,Wu2011}. 
The behavior has been demonstrated in Fig.\,\ref{fig:Over}{(b)} \cite{Grube17}, {which presents the experimental observations} in CeCu$_{6-x}$Au$_x$ across multiple tuning parameters.   
{Tuning the system in the directions that are orthogonal to the gradient of entropy, the distance to the QCP remains unchanged. The gradient of the entropy vanishes precisely at the QCP, which indicates the accumulation of entropy in the quantum critical regime.} 
In this sense, quantum critical systems are particularly soft
and are prone to the formation of unusual excitations and exotic phases.

That strange metals develop via quantum criticality is 
clearly demonstrated in heavy fermion metals. We illustrate the point in YbRh$_2$Si$_2$ and CeRhIn$_5$, via their respective phase diagrams 
{shown in Fig.\,\ref{fig:Over}(c,d)}. The colour coding of $\gamma$ in the figure represents the exponent of the resistivity's dependence on temperature, so regions where $\gamma \simeq 1$ represent the strange metal regime.
Both exhibit an AF order at ambient conditions. In YbRh$_2$Si$_2$, a magnetic field applied perpendicular to
its tetragonal plane of about $0.7$ T (or one applied 
within the plane of about $66$ mT) tunes the system to 
its quantum critical point (QCP) \cite{Geg02.1},
where a $T$-linear resistivity \cite{Bruin13}
occurs over more than three decades in temperature \cite{Ngu21.1}. 
In CeRhIn$_5$, a quantum critical fan develops near a pressure of $2.3$ GPa 
\cite{park-nature06,Kne08.1} with a nearly-$T$-linear resistivity \cite{Par08.1}.

\begin{figure}[h!]
\includegraphics*[width=0.5\textwidth]{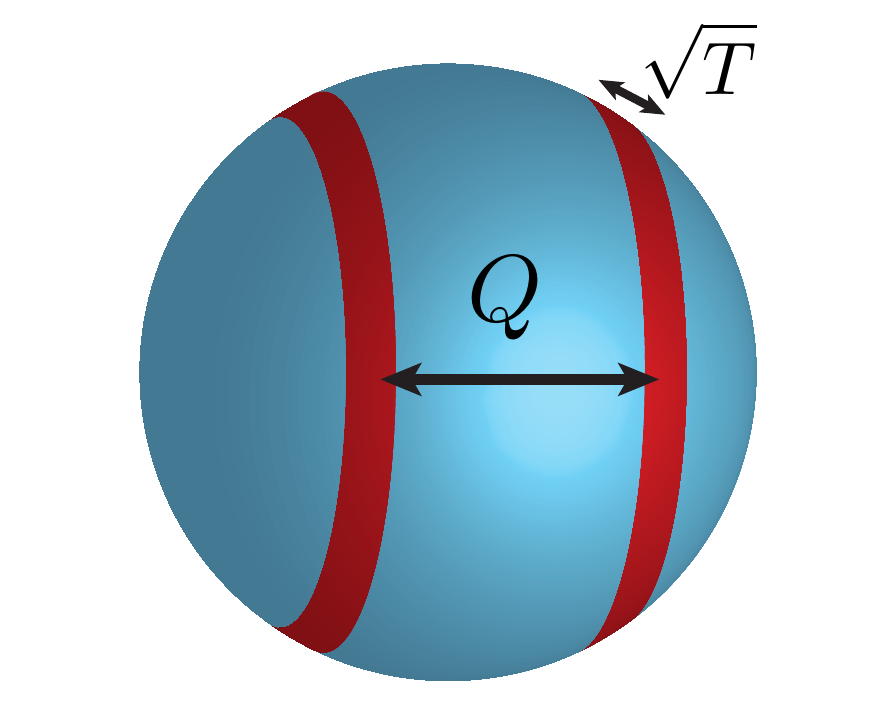}
\caption{\label{fig:FS} {\bf Schematic illustration of the Fermi surface for an SDW QCP.} Schematic illustration of the Fermi surface
for a SDW QCP.
States in
a small portion of the Fermi surface (red stripes with a width of $\sim  \sqrt{T}$ for three dimensions)
can be scattered by the low-energy critical bosons of 
wave vector $Q$. These states are hot in that they experience strong scattering by the order parameter fluctuations.
Meanwhile, the majority of the Fermi surface remains cold (blue region), where Landau quasiparticles are left intact. The electrical transport is dominated by the contributions from the
cold region of the Fermi surface 
and, thus, 
will not show strange metal behavior.}
\end{figure}

Theories of metallic QCPs have two general types. One class of theory is based on the fluctuations of Landau's order parameter,  as described by the Hertz-Millis-Moriya approach \cite{Hertz,Millis}. 
Typically, this order parameter corresponds to a spin-density-wave (SDW) order at an AF wavevector ${\bf Q}$.  
In this case, 
the nonzero ordering wavevector ${\bf Q}$ links narrow hot regions of the Fermi surface to each other. 
The order parameter fluctuations couple to electrons from a small portion (hot region) of the Fermi surface, as shown in Fig. \ref{fig:FS}. Meanwhile, the majority of the Fermi surface remains cold in the sense that
the order parameter fluctuation connects one point on the cold region of the Fermi surface to another point in the Brillouin zone 
where the energy level lies
substantially away from the Fermi energy.
Correspondingly, for the electronic states in the cold region
of the Fermi surface, 
the quantum critical fluctuations have a minimal effect  
and the quasiparticles retain their integrity
\cite{Hlubina95,Rosch99,Borges22.1x}. The electrical transport will not show the strange-metal behavior given that the quasiparticles, 
being long-lived,
will short-circuit the electrical transport.

To realize the strange-metal behavior, 
it is necessary to destroy
the quasiparticles on the entire Fermi surface. This takes place in the second type of theory for metallic quantum criticality,
which goes beyond the Landau framework \cite{Si-Nature,Colemanetal,senthil2004a}.

Here, we survey the beyond-Landau quantum criticality. We start by considering 
how quasiparticles can be critically destroyed. 
The central theme here is that, 
for bad 
metals such as heavy fermion systems, the quasiparticles are fragile to begin with and their formation takes place through a process that is non-perturbative in electron correlations, and yet well-understood. This understanding sets the stage for confronting the 
central challenge, which is how the quasiparticles 
are lost. For heavy fermion metals, the Kondo effect underlies the formation of heavy quasiparticles, whereas the Kondo destruction leads to their suppression. We suggest that these understandings are relevant to the loss of quasiparticles in a variety of strongly correlated systems, including the doped cuprates,
the iron chalcogenides and certain organic superconductors. 
In addition to surveying the theoretical issues, we will 
describe some of the salient experimental developments
\cite{paschen2004,Friedemann.10,shishido2005,Schroder,Aro95.1,Prochaska2020,Maksimovic2022}.

\section{Quantum critical metals -- how to destroy quasiparticles}

To see how the quasiparticles can be lost everywhere on the Fermi surface, we start from their formation away from the QCP.

\subsection{Quasiparticles: the robust and the fragile}

For quantum many-body systems, the physics at low energies is analyzed in terms of building blocks and their 
symmetry-allowed interactions \cite{Pas21.1}. Traditionally, one takes bare electrons as the building blocks and treat the electron-electron interactions order by order in perturbation theory \cite{Nozieres18}. 
The notion of quasiparticles survives up to infinite order of the perturbation series. 
In that sense, quasiparticles are rather robust. For a long time, the validity of Fermi liquid theory was largely unquestioned for systems in dimensions higher than one; indeed, Fermi liquid 
was considered to be the only fixed point of the renormalization-group (RG) flow in such dimensions
\cite{Shankar1994,Polchinski1992}. A quasiparticle
corresponds to a sharp peak in the electron spectral function
as a function of 
energy for a fixed wavevector. 
The wavevectors of zero 
energy excitations form a Fermi surface;
the volume enclosed by the Fermi surface, according to Luttinger's theorem, is proportional to the number of the underlying electrons even in the presence of interactions~\cite{Oshikawa}.
The quasiparticle
has the physical meaning of a dressed electron;
its quantum numbers are exactly 
those of
a bare electron or hole,
namely
charge $\pm e$ and spin $
\frac{\hbar}{2}$. 
Their Fermi statistics dictates a decay rate that goes as $(k_{\rm B}T)^2$, or as $E^2$ as the energy measured from the Fermi energy, $E$, goes to zero. 
In the language of Green's functions,
the self-energy $\Sigma({\km},\omega)$ 
retains the Fermi liquid form up to infinite orders of the perturbative expansion  \cite{Nozieres18}. 
This turns out to ensure a nonzero 
value for the quasiparticle weight, $Z_{\km}$.

Sufficiently strong electron correlations can lead to 
other forms of the building blocks for the low-energy physics. 
For example, heavy fermion systems involve local $f$-electron-derived
moments and itinerant $spd$-electron
bands as the starting point for the description of their low-energy properties \cite{Pas21.1,StewartRMP,Coleman-Nature,Si_Science10,KirchnerRMP}. In that case, quasiparticles are fragile, with a weight that is exponentially small.

Consider the Kondo lattice Hamiltonian, which is described in Box 1.
We start from the parameter regime when the Kondo interaction between the local moments and the itinerant electrons succeeds in driving the formation of a Kondo singlet, which can be pictured as a bound state between a local moment and
a triplet 
particle-hole combination of the conduction electrons. 
Breaking the bound state leads to not only bare conduction electrons, but also a composite heavy fermion formed between the local moment and a conduction electron. The composite fermions have the
same 
quantum numbers 
as bare electrons,
and they hybridize with the conduction electrons 
to form heavy quasiparticles.
These quasiparticles 
have 
a large effective mass and a 
small 
quasiparticle weight 
$Z$ that
is exponentially small and, in practice, 
is of the order $10^{-3}$.

When the quasiparticles are this fragile, competing interactions can readily destroy them.

\subsection{Quantum criticality from Kondo destruction}

The notion of Kondo destruction quantum criticality invokes fluctuations that go beyond a Landau order parameter.
For Kondo lattice systems, it captures the dynamical competition between the Kondo interaction described above and RKKY interactions, which are interactions between the local moments mediated by the spins of the itinerent electrons as described in Box 1.
The 
corresponding QCP is illustrated in Fig.\,\ref{fig:KDQCP}(a)
\cite{Si-Nature,KirchnerRMP}, 
in the space of temperature and non-thermal control
parameter,
$\delta=T_K^0/I$, 
the ratio of the bare Kondo temperature
to the RKKY interaction $I$.

When $\delta$ is sufficiently large,
the Kondo interaction dominates and
a Kondo singlet is formed in the ground state, as illustrated in Fig.\,\ref{fig:KDQCP}(c). As the RKKY interaction is increased, meaning when the parameter $\delta$ is tuned downward, the RKKY interaction 
becomes important and promotes correlations of a spin singlet between the local moments. This process is detrimental to the formation of the Kondo singlet. When it suppresses the Kondo singlet in the ground state, the
composite heavy quasiparticles are 
lost.

Thus, both the formation and loss of quasiparticles can be
considered by analyzing the fate of the Kondo singlet or,
more specifically, 
the amplitude of the Kondo singlet in the ground state. Our strategy is to start from the Kondo side, 
and see whether and how the dynamical competition of the RKKY interaction brings about the suppression of this Kondo-singlet amplitude. One can in principle also 
work from the opposite end, by analyzing the Kondo lattice 
in terms of a quantum nonlinear sigma model representation; the results of such analyses \cite{Si.06,Yamamoto07,Ong_Jones09,Yamamoto-JLTP2010,Goswami-prl11,Goswami-prb14,Goswami-prb17,CCLiu-prb17} are consistent with the conclusions we present here.

Box 1 
provides further details on how the dynamical competition from the RKKY interactions suppresses the Kondo singlet and, by extension, quasiparticles. 
The key is a new fixed point, marked red in Box 1, panel (b).
Here, the Kondo-singlet amplitude vanishes in the ground state,
and the weight of the Landau quasiparticle goes to zero.
This fixed point is 
interacting (as opposed to Gaussian), 
where
$k_{\rm B}T$ is the only energy scale. 

\subsection{Global phase diagram}
\label{subsec:gpd}

The introduction of Kondo destruction  
has inspired considerations of
new quantum phases in the AF Kondo-lattice systems.
These phases
are not only distinguished by the Landau order parameters 
but also by the existence or absence of the Kondo singlet in the ground state.
This has led to a 
global phase diagram, {as given in Fig.\,\ref{fig:GPD}}
 \cite{Si.06,Si_PSSB10,Coleman_Nev}, 
 in the 
 two-parameter space of 
 $J_K$, the Kondo coupling, and
  $G$, which specifies
the extent of the quantum fluctuations in the local-moment magnetism.
The $G$ axis captures the tuning of 
dimensionality \cite{Custers-2012} or
geometrical
frustration
\cite{Zhao-frustration19,Mun12,Kim-2013,Fritsch-2014}.
The quantum phases are distinguished by their magnetic 
behavior and the size of their Fermi surfaces. Here, 
${\rm P}$ and ${\rm AF}$ represent the paramagnetic and antiferromagnetic phases, respectively, while the subscripts ${\rm S}$ and ${\rm L}$ denote small and large Fermi surface, respectively. 
As we explain in Box 1, a large Fermi surface denotes that both the local moments and conduction electrons contribute to the Fermi volume through the Kondo effect, whereas a small Fermi surface signifies the absence of the Kondo effect, with only the conduction electrons contributing to the Fermi volume. 

The stability of the ${\rm AF_S}$ phase
has been analyzed in terms of a quantum nonlinear sigma model
representation of the Kondo lattice
\cite{Si.06,Yamamoto07,Ong_Jones09,Yamamoto-JLTP2010}.
Using the ${\rm AF_S}$ phase as the starting point,
there are three routes for quantum phase transitions to the paramagnetic 
 heavy fermion (${\rm P_L}$) phase.
Trajectory I  describes a direct transition, with a Kondo destruction QCP 
at the border of the AF order.
Trajectory II passes through an intermediate  ${\rm AF_L}$ phase,
which corresponds to the SDW order from the heavy quasiparticles 
of the ${\rm P_L}$ phase.
A Kondo destruction transition takes place inside the AF order,
while the QCP from the AF order to the paramagnetic phase is of the SDW type.
Trajectory III 
passes through
an intermediate ${\rm P_S}$ phase, which could involve non-magnetic order such as a  valence-bond solid or an underlying spin liquid. 
Generically,
the Luttinger theorem of the Kondo lattice is obeyed,
as can be seen from
how the local-moment part and conduction electrons \cite{senthil2004a,Pivovarov.04}
respond to the adiabatic insertion of an external flux \cite{Oshikawa}.
 The 
 paramagnetic heavy fermion (${\rm P_L}$) phase itself, as described earlier,
 represents the standard phase of a Kondo lattice.
 
From the perspective 
of the paramagnetic heavy fermion phase,
the three trajectories of quantum phase transitions 
delineate a variety of ways for the Landau
quasiparticles 
to be destroyed.
Since the initial advancement of the global phase diagram \cite{Si.06}, there has
been considerable effort in exploring this phase diagram, both
theoretically \cite{Yamamoto07,Ong_Jones09,Yamamoto-JLTP2010,Pixley-2014,Goswami-prl11,Goswami-prb14,Goswami-prb17,CCLiu-prb17,Toshihiro2018} and experimentally \cite{Custers-2012,Zhao-frustration19,Mun12,Kim-2013,Fritsch-2014}.
In addition to the Hall effect and quantum oscillations measurements, which we will describe below, thermopower has been utilized to probe the Fermi surface reconstruction and elucidate the global phase diagram \cite{Luo2018}. To illustrate the underlying physics, we will for the most part keep our discussion focused on
the trajectory I of the global phase diagram, which is represented by the phase diagram shown in Fig.\,\ref{fig:KDQCP}(a) in the space of temperature ($T$) and control parameter ($\delta$).
 
 \section{
 Dynamical Planckian scaling}

At the QCP,
$k_B T$ is the only energy scale, and this leads to dynamical properties in which $\hbar \omega$ scales with $k_{\rm B}T$. 
The dynamical spin susceptibility at the AF wavevector $\bQ$ is found \cite{Si-Nature,lcp-prb} to have the following dynamical Planckian scaling form:
\begin{eqnarray}
\chi(\bQ, \omega ) =
\frac{1}{(-i\hbar \omega)^{\alpha}} W^{-1} \left ( \frac{\hbar \omega}{k_{\rm B}T} \right ) \, .
\label{chi-w-T}
\end{eqnarray}
Here, $W=A\,\mathcal{M}(\omega/T)$, with $A$ being a constant prefactor and
\begin{eqnarray}
{\cal M}(\omega/T)
= \left ( \frac{T}{-i\omega} \right )^{\alpha} \exp \,[\,\alpha \,\psi (1/2 - i\omega / 2\pi T)\,] \, ,
\label{cal-M-omega-T-lcp}
\end{eqnarray}
where $\psi$ is the digamma function.

The calculated exponent $\alpha$ is fractional, and is close to being 
 0.75 
 for the Ising anisotropic case 
 (between 0.72 and 0.78 when different methods are used for the calculation)
\cite{GrempelSi,ZhuGrempelSi,Glossop07,Zhu07} and about $0.71$ 
for the case with SU(2) spin symmetry \cite{Hu20.1x}.
At a general wavevector $\bq$, the dynamical spin susceptibility takes the following form:

\begin{eqnarray}
\chi({\bf q}, \omega ) =
\frac{1}{\theta({\bf q}) + A \,(-i\omega)^{\alpha} \mathcal{M}(\omega/T)} \quad .
\label{chi-qw-T}
\end{eqnarray}
Here, $\theta (\bq) = I_{\bQ}-I_{\bq}$,
where $I_{\bq}$ is the RKKY interaction expressed in wave vector space. The comparable critical exponents, obtained from
calculations at the QCPs of the Ising-anisotropic and SU(2)-symmetric Kondo lattice models, 
imply the universal quantum critical behaviors of the dynamical spin susceptibility.

These theoretical results 
provide the understanding of the inelastic neutron scattering data measured in CeCu$_{6-x}$Au$_x$ at its quantum critical concentration $x_c=0.1$ \cite{Schroder} (see also Ref.\,\cite{Aro95.1}). The experiments show not only the 
$\hbar \omega /k_{\rm B}T$
scaling form but also a fractional exponent $\alpha \approx 0.75$.

The Kondo destruction QCP also predicted the temperature dependence of the NMR relaxation rate. When the hyperfine form factor does not have a strong dependence on the wavevector, the NMR relaxation rate $1/T_1$ is determined by the local spin susceptibility, leading to \cite{Si-Nature,lcp-prb}:
\begin{equation}
\frac{1}
{T_1} \propto {\rm constant} \, .
\label{T1-local}
\end{equation}
In contrast, if the hyperfine coupling has a strong ${\bf q}$-dependence leading to a cancellation of the contributions from the dynamical spin susceptibility near the AF wavevector (as in the well-known case of the oxygen-site NMR relaxation rate of the optimally-hole-doped cuprates
\cite{Hammel89,Takigawa91}), the NMR relaxation rate has the following temperature dependence \cite{Si03.1}:
\begin{equation}
\frac{1}{T_1} \propto T^{\alpha} \, .
\label{T1-generic-q}
\end{equation}
The results from the 
silicon-site 
NMR experiments in YbRh$_2$Si$_2$ found the NMR relaxation rate to be strongly dependent on the applied magnetic field \cite{Ishida-prb03,Ishida-prl02}. When combined with the $\mu$SR results, they 
have allowed the extraction of the relaxation rate $1/T_1$ at the quantum critical magnetic field
\cite{Si03.1,Ishida-prb03}, 
and the result is consistent with the prediction of 
Eq.\,(\ref{T1-local}). 
Whereas the measured 
copper-site NMR relaxation rate 
in CeCu$_{6-x}$Au$_x$, at the quantum critical concentration $x_c=0.1$, is compatible with 
the expectation 
of
Eq.\,(\ref{T1-generic-q}) \cite{Si03.1}.

Importantly, charge response, particularly the optical conductivity, has also been found to be critical \cite{Prochaska2020}. This would have been unusual for an SDW QCP, where the singular fluctuations are in the magnetic sector. 
Theoretically,
at the Kondo destruction QCP, the engagement of the Kondo process in the quantum criticality suggests the relevance of the single-particle and charge sectors to quantum criticality
\cite{Cai-charge20.1}.
The corresponding responses, 
including the optical conductivity,
obey 
 dynamical Planckian scaling. 
 Experimental evidence for the involvement of the charge sector in quantum criticality has also been provided in beta-YbAlB$_4$ \cite{Kobayashi2023}.
Further evidence for a singular charge response has come from other theoretical studies 
 \cite{Zhu.04,Pix12.1,Komijani19.1}. 

 Finally, as $k_BT$ 
 is the only energy scale at the QCP, the
 electronic scattering 
 rate
 takes the form
$1/\tau \propto (k_BT)/\hbar$.
With the Umklapp scattering that is generically present in quantum critical metals,
this relationship leads to strange-metal behavior in the temperature dependence of the electrical resistivity.

\section{
Transformation of large-to-small Fermi surface and loss of quasiparticles}

An important characteristic of the Kondo destruction quantum criticality is a transformation of a large to small Fermi surface across the QCP. This turns out to be intimately connected to a loss of quasiparticles everywhere on the Fermi surface at the QCP.

\subsection{Large to small Fermi surface transformation  across the QCP}

 In the paramagnetic phase, the ground state has a nonzero amplitude of the Kondo singlet {, describing the strength of the spin singlet between the local moments and conduction electrons as illustrated in Fig.\,
\ref{fig:KDQCP}(c)}. Correspondingly, {composite-heavy fermion excitations, as described by Fig.\,\ref{fig:KDQCP}(e),} develop in the low-energy single-electron 
spectrum{; as mentioned above,}
the Fermi surface is large
in the sense that it incorporates both the conduction electrons and the Kondo-induced composite fermions,  {which we describe in Box 1.}

On the other side of the QCP, the Kond-singlet amplitude vanishes as illustrated by Fig.\,\ref{fig:KDQCP}(b).  The well-defined composite-fermion excitation is absent, and the single-particle excitations are entirely described by the renormalzied conduction electrons as shown in Fig.\,\ref{fig:KDQCP}(d). This leads to a small Fermi surface as shown in
Fig.\,\ref{fig:KDQCP}(f) and Fig.\,\ref{fig:qp-Z}(a), which  incorporates the conduction electrons only as further described in Box 1.

The jump of the large-to-small Fermi surface across the QCP 
is experimentally testable \cite{Si-Nature,lcp-prb,Colemanetal,senthil2004a,Pepin-prl07}.
Across the field-induced QCP in YbRh$_2$Si$_2$, a remarkable sequence of measurements \cite{paschen2004,Friedemann.10}
have identified a rapid isothermal crossover in the 
Hall coefficient
{(more specifically, the normal Hall coefficient)}. The crossover width extrapolates to zero in the $T=0$ limit. This jump of the Hall coefficient provides evidence for a jump in the Fermi surface across the QCP. Moreover, the location of the crossover maps out a new temperature scale as shown by the solid line in Fig.\,\ref{fig:Over}(c)
\cite{paschen2004,Friedemann.10,Gegenwart2007}.
Separately, in CeRhIn$_5$, measurements of the de Haas-van Alphen effect have provided evidence of a sharp jump of the Fermi surface across the pressure-induced QCP \cite{shishido2005}.
Additional evidence for a Fermi-surface transformation across the QCP has come from Hall effect measurements in pressurized CeRhIn$_5$ \cite{WangPark2023}.

\subsection{Loss of quasiparticles at the QCP}

As the system approaches the Kondo destruction QCP from the side of large Fermi surface, the Kondo-singlet amplitude goes to zero. The residue of the pole
in the conduction-electron self-energy 
$\Sigma({\bf k},\omega)$ of Eq.\,(\ref{sigma-pole}) {in Box 1} vanishes. 
Correspondingly, the quasiparticle weight on the large Fermi-surface, $Z_{\rm L}$ vanishes {as illustrated in Fig.\,\ref{fig:qp-Z}(a)}.
Continuity dictates that
the quasiparticle weight
on the small Fermi surface, $Z_{\rm S}$, vanishes 
as well
upon approaching the QCP from the other side of the phase diagram, {as shown in Fig.\,\ref{fig:qp-Z}(a)}.

Direct spectral evidence of the destruction of quasiparticles at the QCP is hard to 
{obtain} because in heavy fermion metals, angle-resolved photoemission spectroscopy (ARPES) measurements have yet to reach adequate resolution to address the issue.
However, scanning tunneling spectroscopy 
in the heavy fermion compound YbRh$_2$Si$_2$ \cite{Ernst2011}
has provided evidence that the single-particle excitations are a part of its quantum criticality \cite{Seiro2018}.
Related evidence has come from the probe of Kondo-driven excitations
in the quantum critical regime by 
time-resolved terahertz 
spectroscopy \cite{Yang-Kroha2023}.
Recently, the current shot noise
has been used as a new probe of strongly correlated metals. The observed reduction of the Fano factor provides fairly direct evidence for the loss of quasiparticles 
in the quantum critical regime of YbRh$_2$Si$_2$ \cite{Liyang-Chen2023,Wang-shot-noise-2022}.

\subsection{Implications of the singular charge response: dynamical Kondo effect and high-$T_{\rm c}$ superconductivity}

That charge responses are singular and obey dynamical Planckian scaling at a magnetic QCP carries a special significance. It implicates a charge-spin entanglement 
at the Kondo destruction QCP, in spite of a vanishing amplitude of the Kondo singlet in the ground state.
In fact, it has been shown that a dynamical Kondo correlation
persists in this regime: a nonzero Kondo coupling in the 
Hamiltonian dictates that the cross local moment-conduction electron spin correlations operate at 
nonzero frequencies
\cite{Hu20.1x}. 
More generally, recent work, both theoretical~\cite{Fang2024_QFI} and experimental~\cite{Silke2024_QFI}, has provided 
evidence for amplified entanglement at the Kondo-destruction QCP.

Qualitatively, the Kondo destruction QCP{, described in Figs.\,\ref{fig:KDQCP} and \ref{fig:qp-Z}}, features quantum fluctuations between a phase 
that has a Kondo singlet in the ground state
and with an accompanying large Fermi surface on the one hand, and a phase that has no
Kondo singlet in the ground state
and with a corresponding
small Fermi surface on the other hand. 
Since the composite fermions carry both charge and spin, the fact that they are critically 
suppressed at the Kondo destruction QCP means that the charge sector is an inherent component of the quantum criticality{, as shown in 
Fig.\,\ref{fig:qp-Z}(b)}.

This dynamical Kondo effect has important implications for how unconventional superconductivity develops out of the strange-metal normal state. The Kondo destruction quantum criticality is robust in that 
a large 
entropy 
-- amounting to a significant portion of $R$$\ln2$, with $R$ the ideal gas constant,  per $f-$site --is encoded
in the quantum fluctuation spectrum.
The 
primary degrees of freedom that is involved in this amplified quantum fluctuations are spin in nature. Indeed, a recent calculation using the cluster version of the 
{extended dynamical mean-field theory (EDMFT)} 
found large intersite spin-singlet correlations in this quantum critical fluid \cite{Hu2021.sc-x}. Through the dynamical Kondo effect, such amplified quantum fluctuations strongly influence the charge sector.
In turn, the 
singlet spin correlations lead
to pronounced spin-singlet pairing correlations.
The calculations show that this process drives unconventional superconductivity with high-$T_{\rm c}$: the transition temperature reaches a few percent of the effective Fermi temperature \cite{Hu2021.sc-x}.

\section{Implications and broader contexts}

\subsection{Delocalization-localization transition in other 
correlated systems}

We have emphasized how Kondo destruction corresponds to a delocalization-localization transition 
of the $f$-electrons across the QCP.
Such an effect also appears in more complex $f$-electron systems, which involve  entwined local degrees of freedom of both spins and orbitals \cite{Martelli19,Liu_pnas23}.
Localization-delocalization transitions of this kind in a metallic environment 
is emerging as a unifying theme across the correlated material classes.

\begin{figure}[b!]
\includegraphics*[width=
1.0\textwidth]{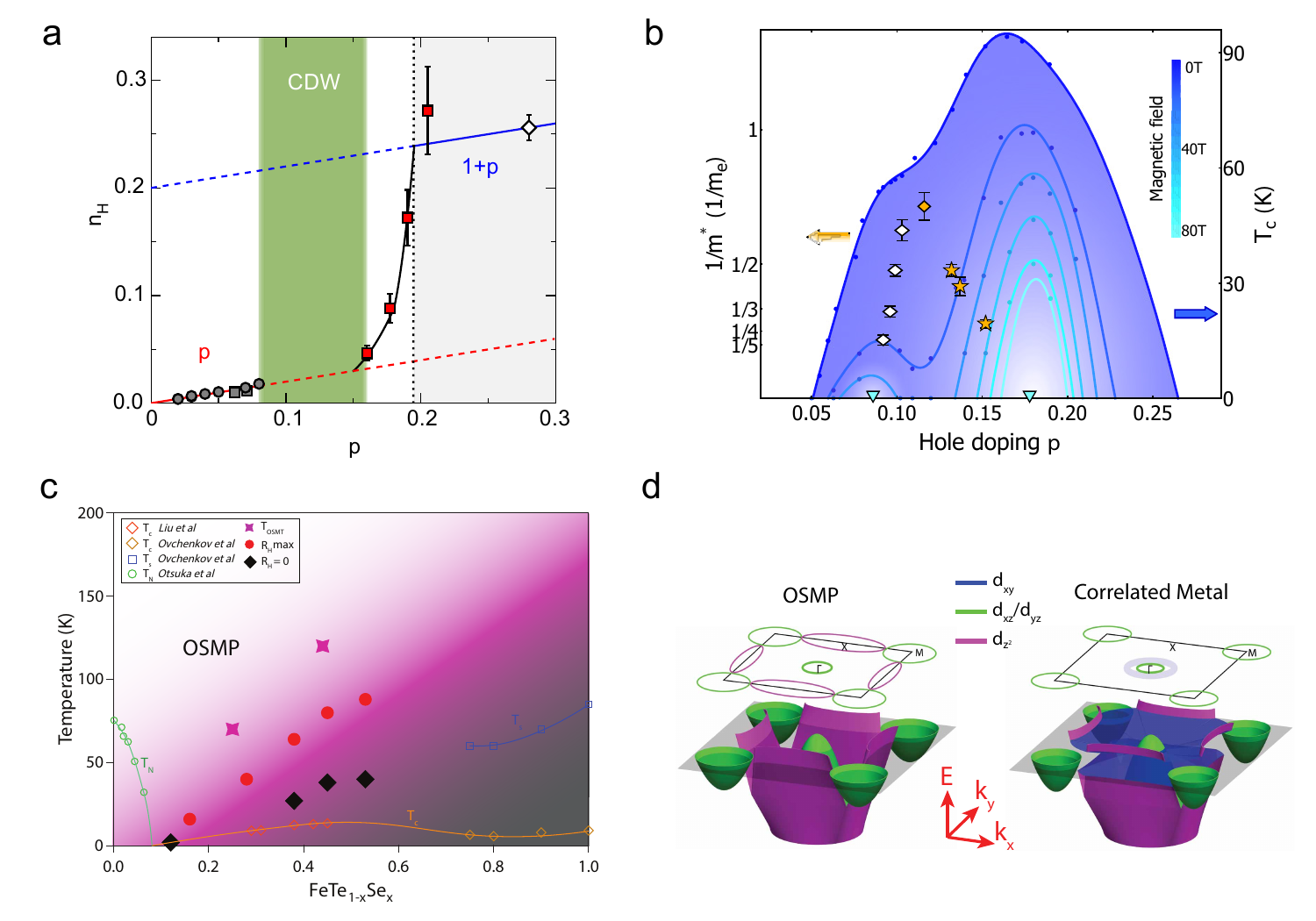}



\caption{\label{fig:QCP} {\bf Evidence for localization–delocalization transitions in the cuprates and iron chalcogenides.} {\bf a} Doping ($p$) dependence of the Hall number ($n_H$) in the hole-doped cuprates \cite{Bad16.1}. The red and blue lines correspond to $n_H=p$ and $n_H=1+p$, respectively. {\bf b} Effective mass $m^*$ enhancement near the optimally doped YBCO under a high magnetic field, with the blue curves denoting 
$T_c$ at different magnetic fields
\cite{Ram15.1}.  {\bf c} Phase diagram of FeTe$_{1-x}$Se$_{x}$, with $T_{N}$, $T_{s}$ and $T_c$ respectively representing the
temperatures for AF, structural and  superconducting transitions. 
The red dots and black diamonds correspond to the temperatures for 
the maximum ($R_H\text{max}$) and zero ($R_H=0$) 
of Hall resistivity.
The orbital-selective Mott transition temperature
($T_{\rm OSMT}$) is defined as the temperature 
at which the photoemission spectral weight 
of the $d_{xy}$ orbital vanishes. The gradual suppression of
$T_{\rm OMST}$ with decreasing $x$ 
provides evidence for an 
orbital selective Mott 
quantum phase 
transition.
{\bf d} Schematic illustration of a large-to-small Fermi surface transformation as $x$ is decreased in FeTe$_{1-x}$Se$_{x}$,
induced by the de-hybridization of the $d_{xy}$ 
orbital associated with the
OSMT \cite{Huang2022}.  }
\end{figure}

In the hole doped cuprates, strange-metal behavior is well established 
\cite{Legros19.1,Giraldo-Gallo2018,Tak92.1,Martin-Tlinear-prb90,Varma_rmp2020,Phillips-science22}.
Hall effect measurements in 
YBa$_2$Cu$_3$O$_y$ (YBCO), when combined with the results in 
underdoped La$_{2-x}$Sr$_x$CuO$_4$ (LSCO) and 
overdoped Tl$_2$Ba$_2$CuO$_{6+\delta}$
have implicated 
a transition between phases with 
carrier concentrations $p$ and {$1+p$} near the optimal
doping [Fig.\,\ref{fig:QCP}(a)] \cite{Bad16.1}.
This is accompanied by the observation of mass enhancement 
in YBCO near optimal doping [Fig.\,\ref{fig:QCP}(b)] \cite{Ram15.1}. While this remains an issue of 
active discussions \cite{Putzke-natphys21,Bal03.1}, 
the notion that the Fermi surface undergoes 
a small-to-large transformation as a function 
of hope doing has also been reported based on 
angle-dependent magnetoresistance  measurements in 
La$_{1.6-x}$Nd$_{0.4}$Sr$_{x}$CuO$_4$ \cite{Fang2022}.
Moreover, recent inelastic measurements 
have implicated a QCP near the optimal doping in LSCO \cite{ZhuHayden2023}.
All these provide evidence for the relevance of a QCP involving an
electronic localization-delocalization to the physics of optimally hole-doped cuprates.

A localization-delocalization transition has also been evidenced in other correlated electron material classes. 
An orbital-selective Mott transition, with a large-to-small Fermi surface transformation, has been 
{demonstrated by} FeTe$_{1-x}$Se$_x$ by ARPES measurements {as shown in Fig.\,\ref{fig:QCP}(c,d)} \cite{Huang2022}.
In a doped Mott insulator of organic charge-transfer salts,
some preliminary evidence for a rapid Fermi surface change in a doped Mott insulator has also emerged from Hall \cite{Oik15.1}
and thermoelectric \cite{Wakamatsu22.1x} measurements,
In moir\'{e} systems, related properties are also being
uncovered 
\cite{Cao18.1,Jao-StrangeM22.1}. 
Intriguingly, $T_c/T_0$ of the moir\'{e} systems is also on the order of a few percent \cite{Cao18.1}.
Finally, kagome and related metals 
with frustrated lattices, with active flat bands, have recently emerged as a platform for strange metal behavior \cite{Hua23.2x,Ye21.1x,Eka22.1x}.
A Kondo lattice description 
\cite{Hu22.3x,Che22.3x}
allows for strange metallicity 
in terms of localization-delocaliation of electrons in compact molecular orbitals \cite{Che23.1x}; this approach leads to
a phase diagram of temperature and control parameter that has now been supported by
an experimentally determined temperature-pressure phase diagram in a kagome metal \cite{Liu2023.x}. 



\subsection{Quantum critical point vs. quantum critical phase}

\begin{figure}[h!]
\includegraphics*[width=
1.0\textwidth]{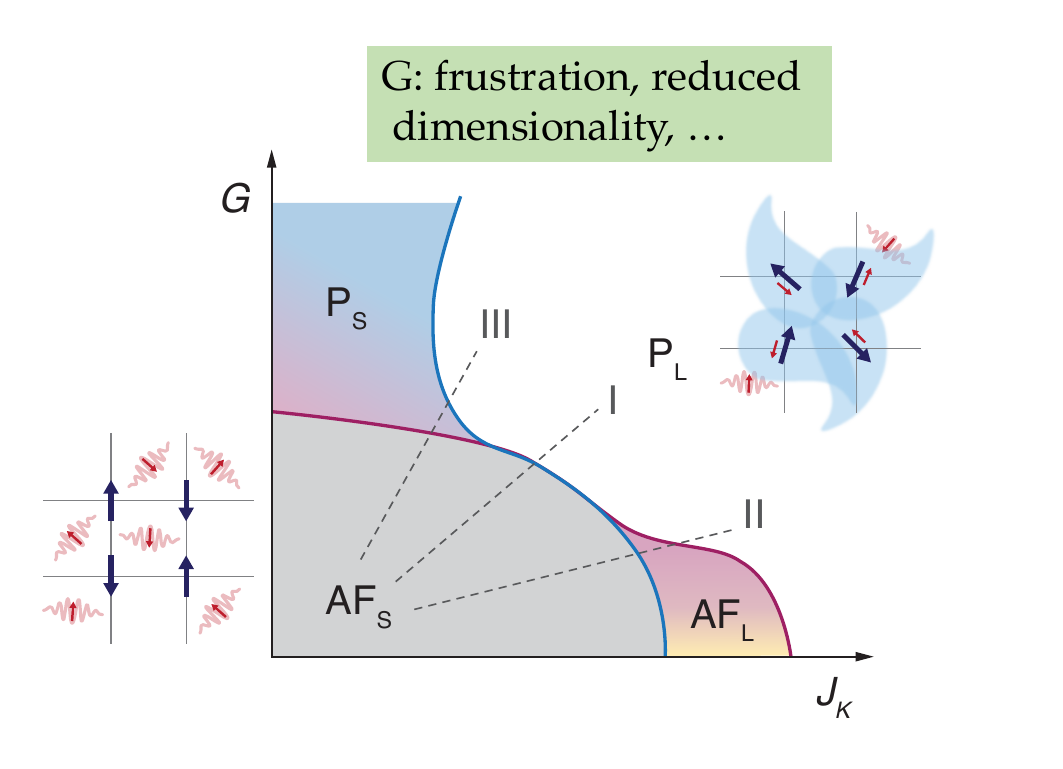}
\caption{\label{fig:GPD} {\bf Global phase diagram for quantum critical phases and points from amplified quantum fluctuations.} Global phase diagram of the heavy fermion systems in the parameter space of $J_K$, 
the Kondo coupling,
and $G$, which characterizes the quantum fluctuations in the local-moment magnetism. 
Adapted from Refs.\,\cite{Si.06,Si_PSSB10,Pixley-2014,Pas21.1}.  
}
\end{figure}

One of the topical issues in the realm of quantum criticality concerns the possibility of a quantum critical phase. The global phase diagram for quantum critical heavy fermion metals,
described earlier {in} Sec.\,\ref{subsec:gpd} and Fig.\,\ref{fig:GPD},
delineates the close relationship between the two possibilities. Here, a quantum critical phase can develop in the regime
${\rm P}_{\rm S}$, where quantum fluctuations prevent the system from acquiring a long-range order.
The global phase diagram suggests that both the quantum critical phase and the beyond-Landau Kondo destruction quantum critical points descend from the same phenomenon, namely the strong dynamical competition between the Kondo and RKKY interactions. 
In a heavy fermion compound with 
geometrical frustration (due to a distorted kagome lattice), CePdAl, a quantum critical phase has been implicated in its pressure-magnetic field phase diagram \cite{Zhao-frustration19}. Evidence for a quantum critical phase has also come from 
thermoelectric measurements in
an organic charge-transfer salt \cite{Wakamatsu22.1x}.

In the cuprates,
we have already discussed evidence for 
the relevance of a quantum critical point 
\cite{Bad16.1,Ram15.1,Fang2022,ZhuHayden2023}.
In the LSCO family,
the evidence for a quantum critical point developing near optimal superconductivity
includes the observation of a peak in the specific heat (and, correspondingly,
the maximization of entropy),
and the presence of low-energy collective spin fluctuations with an energy scale comparable to temperature
\cite{ZhuHayden2023}. 
On the other hand, experimental observations have revealed that both the linear-$T$ behavior in the resistivity for LSCO
\cite{Cooper-Hussey2009,Ayres2021}
and a quadrature scaling in the magnetoresistance for 
Tl$_2$Ba$_2$CuO$_{6+\delta}$
and Bi$_2$Sr$_2$CuO$_{6+\delta}$ 
\cite{Ayres2021} occur at doping levels beyond $p^*$, 
raising the possibility
of a quantum critical phase. 
Regardless of whether 
the physics is driven by a quantum critical point or a quantum critical phase,
the phenomenology suggests that both 
collective spin fluctuations and the 
electron localization-delocalization transition 
are involved in the low-energy physics
in the strange metal 
regime. This bears
similarity to the phenomena observed in heavy-fermion systems.



\section{Summary and outlook}

We have highlighted the theme that quasiparticles are fragile to begin with in strongly correlated metals such as 
heavy fermion systems and that, in the Kondo destruction quantum criticality, the quasiparticles are lost at the 
delocalization-localization transition of the $f$-electrons. This theme unveils a hidden Mott transition in an unlikely
setting, namely between two metallic phases.  As such, the unusualness of the properties here rivals what happens
in the case of the standard Mott transition. 
By certain measure, it is even more striking 
because, with both sides of the 
transition being metallic, the 
Coulomb interactions are screened and it is more natural to have the quantum phase transition 
to be continuous. The loss of quasiparticles at the Kondo destruction QCP is accompanied not only by 
such spectacular feature as dynamical Planckian scaling in the 
spin and charge dynamics but also in a sudden 
transformation between large and small Fermi surfaces across the QCP. 

These salient properties allow 
one to connect the strange metallicity of heavy fermion metals with that of a variety of strongly
correlated systems. The strange metallicity in the cuprates and organic systems 
naturally develop in the backdrop of the parent Mott insulator phase. 
In the iron-based superconductors, recent experiments have provided evidence for the proximate orbital-selective 
Mott phase. Finally, in moir\'{e} and frustrated lattice systems, where strange metal behavior has also been observed, correlated insulating 
phases may well be considered as the result of electron localization. It appears to be no coincidence that 
the strange metal behavior develops in all these strongly correlated material classes and that superconductivity emerges with a high
transition temperature.
 By extension, it seems likely that, in most if not all
of these systems, strange metallicity
is underlined by the loss of quasiparticles on the entire Fermi surface. Exploring the issues
in 
diverse settings and 
from varied perspectives  
promises to deepen the understanding about quantum critical metals and to uncover new connections that 
the strange metal physics of heavy fermion metals may have 
with 
that
of 
a broad range of
other
correlated material classes.



\acknowledgements

We would like to thank  A. Cai, J. Cano, L. Y. Chen, L. Deng, Y. Fang, M. T. Glossop, P. Goswami,
 the late D. R. Grempel, K. Grube, N. E. Hussey, K. Ingersent, A. Kandala, S. Kirchner, J. Kono, X.-W. Li, C.-C. Liu, M. B. Maple, D. Natelson, E. M. Nica, C. P\'epin, J. H. Pixley, L. Prochaska, C. Setty, J. L. Smith, O. Stockert, S. Sur, J. D. Thompson, M. G. Vergniory, H. von L\"{o}hneysen, Y. Wang, S. Wirth, J.-D. Wu, R. Yu, F. Xie, 
 J.-X. Zhu, L. J. Zhu and,
 particularly S. Paschen and F. Steglich,
 for collaborations 
 and discussions.
We are grateful to J. D. Thompson for supplying Fig.\,\ref{fig:Over}(a).
This work has been supported in part by
the NSF Grant No.\ DMR-2220603 (H.H.), 
the AFOSR under 
Grant No.\ FA9550-21-1-0356 (L.C.),
the Robert A. Welch Foundation Grant No.\ C-1411 (Q.S.), 
the Vannevar Bush Faculty Fellowship ONR-VB N00014-23-1-2870 (Q.S.),
and the 
European Research Council (ERC) under the European Union’s Horizon 2020 research and innovation program (Grant Agreement No. 101020833) (H.H). 
\\
\\
\noindent{\bf Author contributions:}~~
H.H., L.C. and Q.S. contributed to the 
writing of the article.

\vspace{0.2cm}
{\noindent{\bf Competing 
 interests:}~~
The authors declare no competing 
 interests. }

\clearpage
\newpage

\clearpage
\newpage

\begin{figure}[h!]
\vspace{-1cm}
\includegraphics*[width=1.0\textwidth]{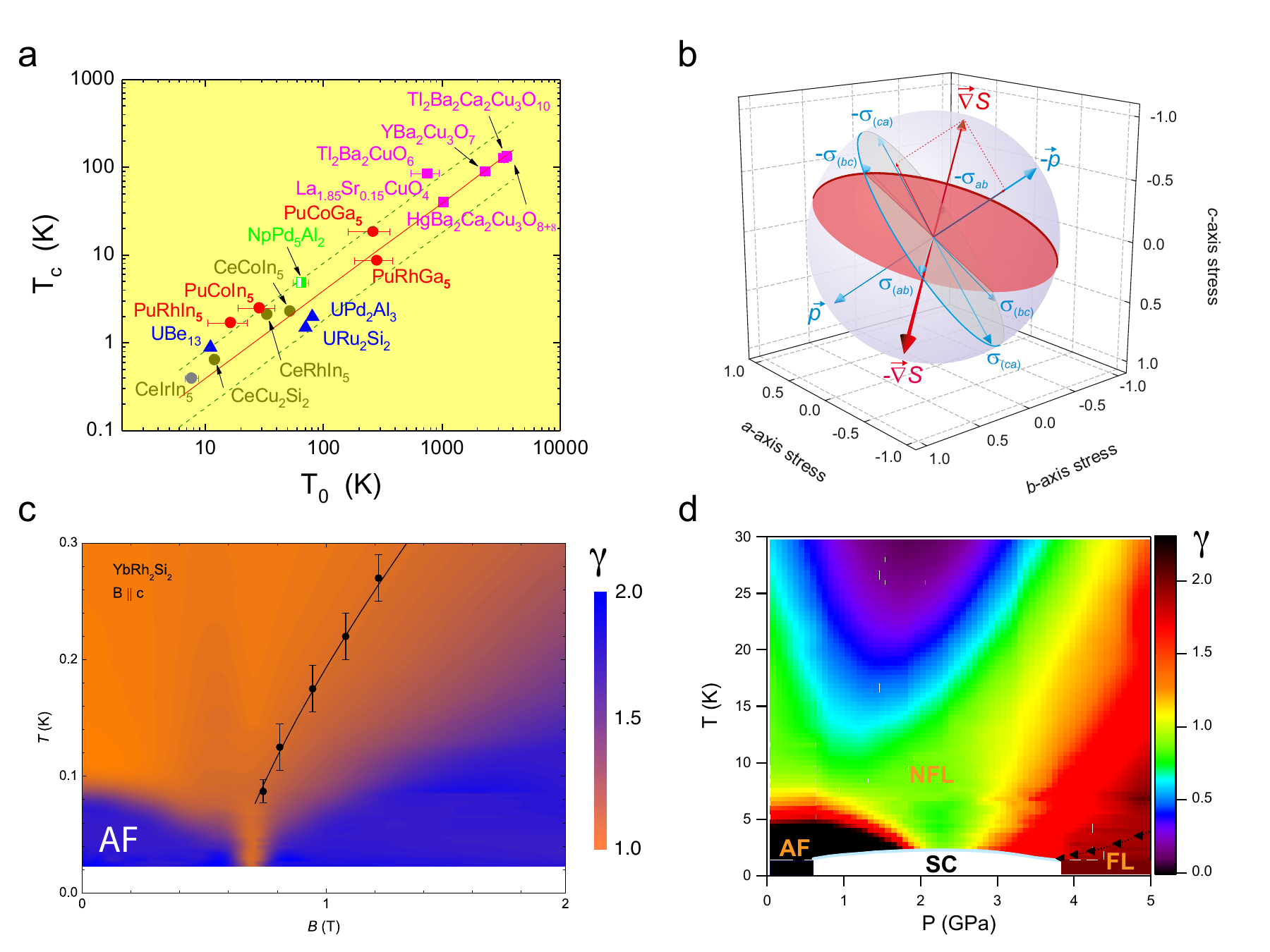}

\caption{\label{fig:Over} {\bf  Strange metallicity and superconductivity.} 
{\bf a,} Superconducting transition temperature ($T_{c}$) versus the effective Fermi temperature ($T_{0}$, extracted from entropy and other means) for various superconductors from different material families (adapted from Ref.\,\cite{Curro05}, 
courtesy J. D. Thompson). The red solid line indicates the linear proportionality between the two temperature scales. All materials, marked by the different colours and symbols, approximately follow this behaviour and are located within the region bounded by the two green dashed lines. The error bars reflect the uncertainties in the measurements and analyses. {\bf b,} Illustration of the evolution in the entropy near a QCP, based on measurements of the uniaxial Gr\"uneisen ratios in CeCu$_{6-x}$Au$_x$ at its critical concentration $x_c =0.1$. Shown here is the parameter space of the uniaxial stress ($p$) along the crystalline $a$, $b$, $c$ axis. $\sigma(ab)$ represents the shear pressure. The red arrow marks the direction with the steepest slope of entropy $\nabla S$, with the directions perpendicular to it marked by the red plane. 
 }
\end{figure}
\begin{figure}[t]
  \contcaption{
 (cont'd) {\bf c,d,} Temperature-control parameter phase diagrams of YbRh$_2$Si$_2$ under a magnetic field ($B$) applied along the crystalline $c$ direction of the system~\cite{Geg02.1,Custers2003} ({\bf c}) and CeRhIn$_5$ under pressure ($P$)~\cite{Par08.1} ({\bf d}). The colours represent the temperature exponent $\gamma$ of the resistivity ($\rho$), determined by a logarithmic derivative of $\Delta\rho(T) \equiv \rho(T) - \rho(T= 0)$ with respect to ${\rm log}T$, signifying a $T^{\gamma}$ dependence. In panel {\bf c}, the black solid line tracks the evolution in the Hall coefficient and thermodynamic quantities that indicate a crossover between large and small Fermi surfaces. The error bars reflect the uncertainty of the extracted crossover location. In panel {\bf d}, the phase boundaries are displayed for the local-moment antiferromagnetic (AF) order, the superconducting (SC) phase, and the regions (below the black dotted line) where the resistivity follows a $T^2$ temperature dependence, characteristic of a Landau Fermi liquid (FL). The cone-shaped green region denotes the non-Fermi liquid (NFL) regime with a sub-linear $T$-dependence in the resistivity. }
\end{figure}

\clearpage



\begin{figure}[h!]
\includegraphics*[width=0.8\textwidth]{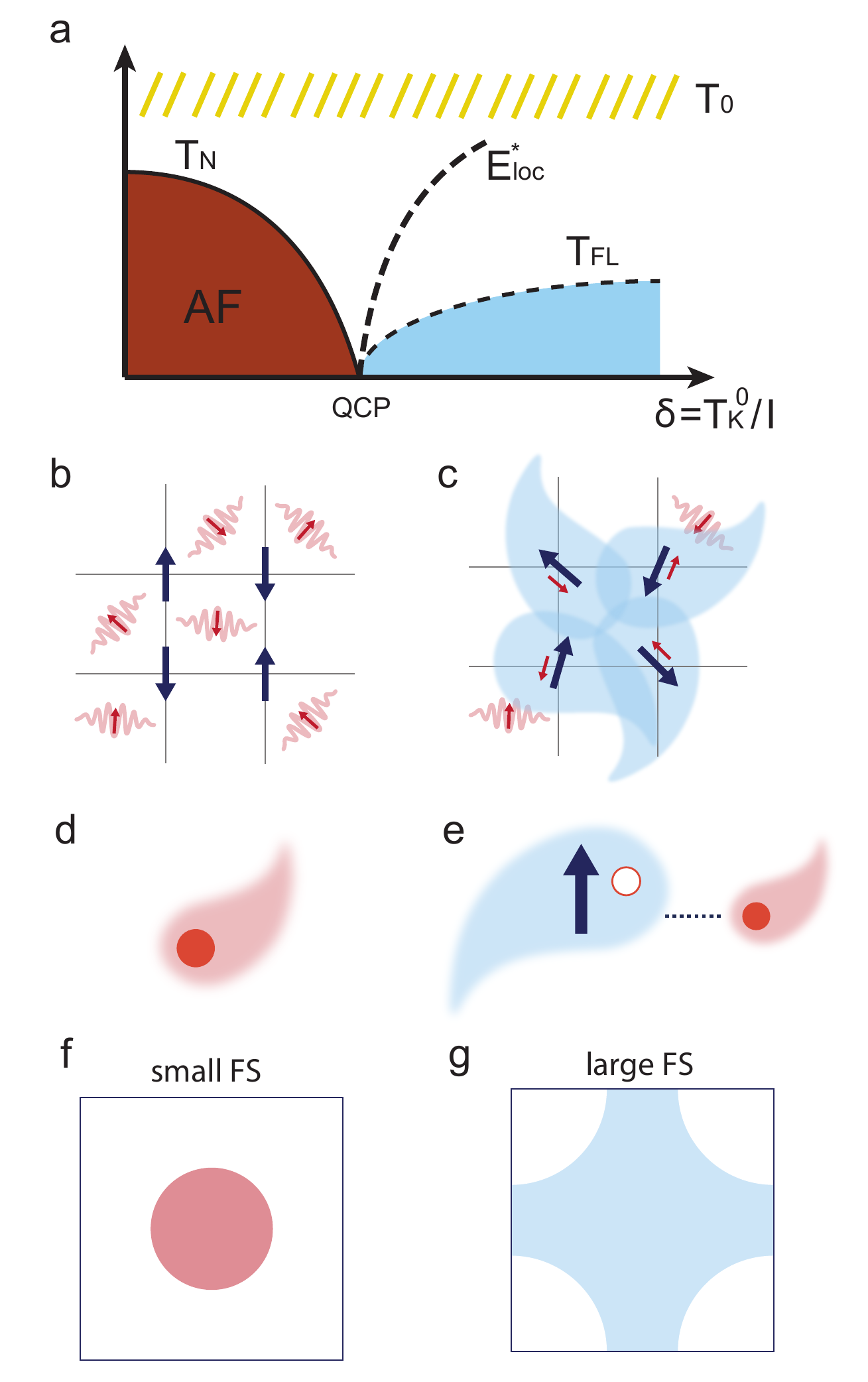}

\caption{\label{fig:KDQCP} {\bf Kondo destruction quantum criticality and large-to-small Fermi surface transformation.} {\bf a} Kondo destruction quantum criticality of a Kondo lattice, 
}
\end{figure}
\begin{figure}[t]
  \contcaption{(cont'd) in which 
a Kondo destruction energy scale 
$E_{\rm loc}^{*}$ vanishes 
at the QCP. Here, $T_{\rm N}$, $T_{\rm FL}$ and $T_0$ denote the temperatures for an AF ordering, the crossover into a
Fermi liquid state and the initial onset of Kondo correlations, respectively. On the two sides of the QCP 
are {\bf b} an AF
order without the formation of a Kondo singlet
and {\bf c} a paramagnetic phase with a Kondo-singlet ground state. 
In the AF phase, the quasiparticles only involve {\bf d} the conduction electrons, in contrast to the paramagnetic phase, in which {\bf e} the Kondo singlets in the ground state yield composite heavy fermions in the excitation spectrum. 
The Fermi surface for the AF phase is small {\bf f} 
in that it only involves the conduction electrons, while that for the paramagnetic phase is large {\bf g} in that it also counts the number of local moments. Panel {\bf a} and {\bf b-g} are adapted from Refs.\,\cite{Si-Nature,KirchnerRMP} and
Ref.\,\cite{Pas21.1}, respectively.
}
\end{figure}

\clearpage 
\newpage 

\begin{figure}[h!]
\includegraphics*[width=
0.7\textwidth]{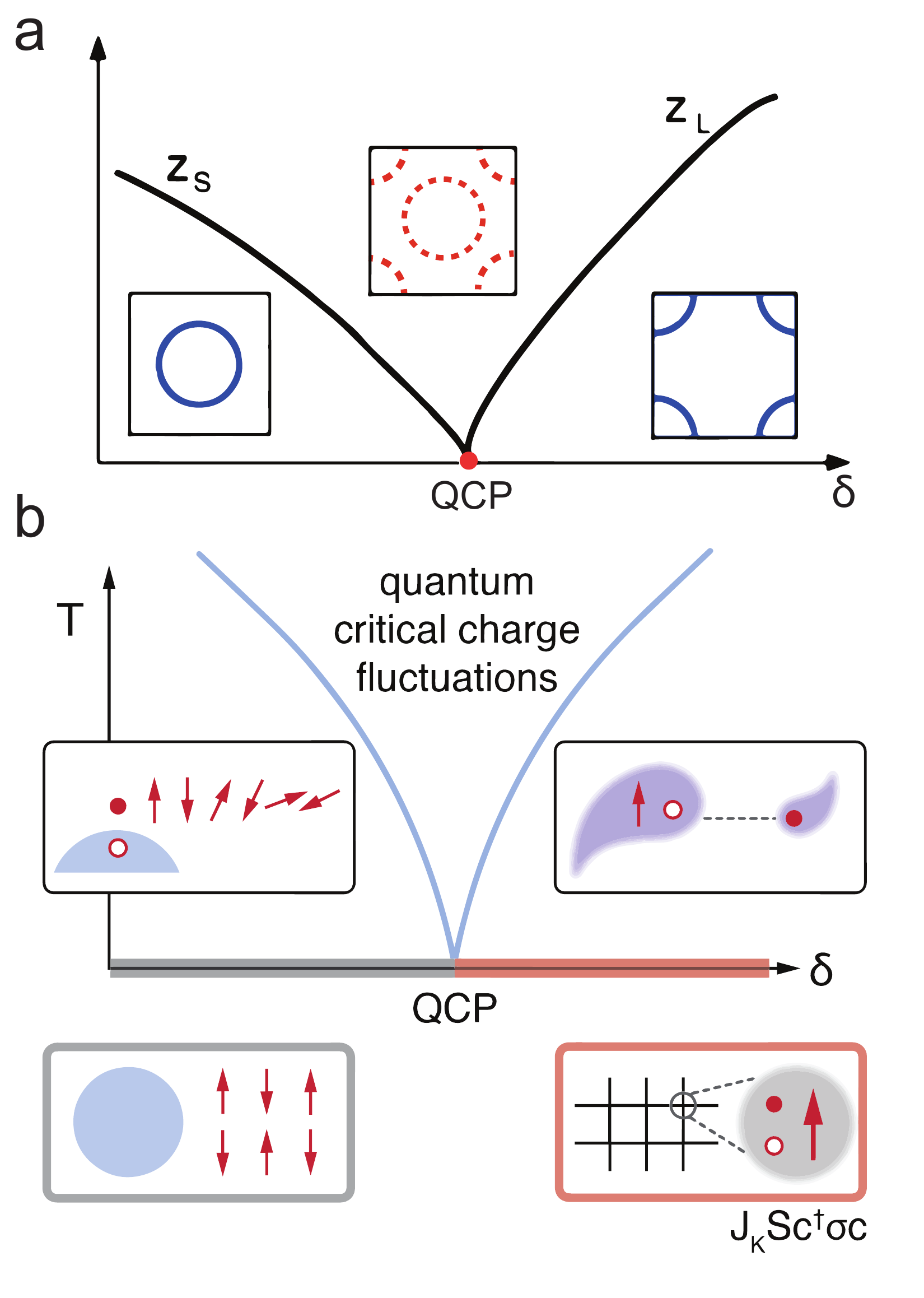}





\caption{
\label{fig:qp-Z} {\bf Loss of quasiparticles and quantum critical charge fluctuations.} {\bf a} The quasiparticle spectral weight, $Z_S$ and $Z_L$, for the small and large Fermi surfaces respectively as a 
function of $\delta$, showing a large-to-small Fermi surface transformation across the Kondo destruction QCP, and the loss of quasiparticles everywhere on the Fermi surface at the QCP \cite{KirchnerRMP}.
{\bf b} Illustration of the quantum critical charge fluctuation at the Kondo destruction QCP 
between an 
AF state, }
\end{figure} 
\begin{figure}[t]
  \contcaption{(cont'd) illustrated by the bottom left box where 
the staggered red arrows denote the long-range magnetic order of local moments and the blue circles represent the Fermi surface of the conduction electrons, and a paramagnetic state,  
  illustrated by the bottom right box where a Kondo singlet is formed 
 between the local moments (red arrow) and the spins of 
 conduction electrons in the form of a spin-triplet combination of particles and holes
  (red solid and open dots).
Adapted from Ref.\,\cite{Prochaska2020}.}
\end{figure}

\clearpage
\newpage

\begin{InfoBox}[h]
\caption{{\bf Kondo lattice system 
}\label{box_kondo-lattice}} 
\fbox{
\begin{minipage}{0.98\textwidth}\raggedright
Consider the Kondo lattice Hamiltonian:
\begin{eqnarray}\label{eq:KLHamiltonian}
H_{\rm KL}=
\sum_{\km\sigma } \varepsilon_{\km} c_{\km \sigma }^{\dagger} c_{\km \sigma }
+
 \sum_{ ij } I_{ij}
{\bf S}_i \cdot {\bf S}_j
+  \sum_{i} J_K {\bf S}_i \cdot c^{\dagger}_{i}
\frac{\vec{\sigma}}{2} c^{\phantom\dagger}_{i} .
\label{eq:kondo-lattice-model-SU2}
\end{eqnarray}
The involved building blocks are the
$f$-electrons in the form of local moments,
$\bf{S}_i$, and a band of $spd$ conduction electrons,
$c_{\km \sigma}$ with an energy dispersion $\varepsilon^{\phantom\dagger}_{\km}$.
At each site $i$, 
an AF Kondo interaction $J_K$
couples the spin of the local moment and that of 
the conduction
electrons, ${\bf s}_{c,i} = (1/2) c_{i}^{\dagger} 
\vec{\sigma} c_i$,
where $\vec{\sigma}$ denote the three Pauli matrices.
Across the sites, the local moments are coupled to each other via an RKKY interaction $I_{ij}$.

The calculations that have provided the basis for the notion of Kondo destruction
is an 
{EDMFT} \cite{Si.96,SmithSi-edmft,Chitra2001};
for a recent review, see
Ref.\,\cite{Hu-EDMFT-review2022}. This approach treats the dynamical interplay between the Kondo and RKKY interactions of 
a Kondo lattice {described by panel (a)}.
The EDMFT approach corresponds to a non-perturbative summation of an infinite series of skeleton diagrams. They are generated by an effective action functional and are systematic and conserving. 

In this approach, the fate of the Kondo-singlet amplitude is 
characterized by the nature of local correlation functions. The latter are determined from a Bose-Fermi Kondo/Anderson model:
\begin{eqnarray}
H_{BFK} &= &\sum_{ \bk,\sigma }E_{\bk}
c^{\dagger}_{\bk\sigma}
c^{\phantom\dagger}_{\bk\sigma} 
+ \sum_{\bp} \omega_{\bp} \bm{\phi}_{\bp}^{\dag}\cdot 
\bm{\phi}_{\bp}
\nonumber 
\\
&& + J_K \bm{S} \cdot \frac{c_0^\dag \bm{\sigma} c_0}{2} + g :\bm{S}: \cdot \sum_{\bp}( \bm{\phi}_{\bp}+ \bm{\phi}^{\dag}_{-\bp}) 
+ h_{\rm {loc}} S^z \, .
\end{eqnarray}
Here the dispersion $E_{\bk}$ and $\omega_\bp$ 
are associated with a fermionic and a bosonic bath
characterized by a fermionic field $c_{\bk\sigma}$, at momentum $\bk$ and spin $\sigma$, and a bosonic field 
$\bm{\phi}_{\bp}$, at momentum $\bp$. Their couplings to the local moment have the strength of $J_K$ and $g$, respectively. In addition, $h_{\rm loc}$ denotes an effective static magnetic field, which is spontaneously generated and is coupled to the $z$-component of the local spin.
Self-consistency equations are expressed in terms of the local correlators of the Bose-Fermi Kondo model.

The Kondo destruction
is seen from the RG flow of the Bose-Fermi Kondo model. The RG 
is analyzed at one loop from an $\epsilon$-expansion,
first carried out in the model with Ising anisotropy \cite{Si.96} and subsequently extended to the model with SU(2) spin symmetry \cite{SmithSi_EPL1999,Sengupta}. 

\end{minipage}
}
\end{InfoBox}

\clearpage
\newpage
\fbox{
\begin{minipage}{0.98\textwidth}
For 
the SU(2) and $xy$-spin symmetry cases, the RG analysis has 
been carried out to two and higher loops \cite{ZhuSi,ZarandDemler}.
Here, $\epsilon$ 
describes the power-law spectrum of the bosonic bath, which also contains a high-energy cutoff $\Lambda$:
\begin{eqnarray}
\rho_b(\omega) \equiv \sum_{\bp}  \delta (\omega - w _\bp)  
\propto 
|\omega|^{1-\epsilon} ~~~~~~~~\text{for} ~ |\omega| < \Lambda \, .
\label{dos-boson}
\end{eqnarray}
Panel (b) illustrates the flow that is associated with the RG beta-functions in the $\epsilon$-expansion, for a positive $\epsilon$. 
{
The RG flow diagram illustrates two categories of stable fixed points: one associated with an 
infinite $J_K$, representing the Kondo screened phases, and the other characterized by $J_K=0$, indicating Kondo-destroyed
phases. 
An 
unstable fixed 
point
corresponds to 
a Kondo-destruction quantum critical point.}
In the absence of the bosonic Kondo coupling (when $g=0$), any nonzero $J_K$ flows away from the decoupled fixed point and towards the Kondo fixed point \cite{Hewson}. The bosonic coupling $g$ leads to two new fixed points and a separatrix in the $J_K$-$g$ plane. The critical (red) fixed point controls the physics on the separatrix, corresponding to a critical destruction of the Kondo phase. On the right of the separatrix, the system flows to a Kondo-destroyed (green) fixed point 
where $J_K$ vanishes altogether.

The nature of the critical (red) Kondo destruction fixed point is to be contrasted with that of the Kondo fixed point. 
{
The Kondo fixed point is characterized by a nonzero Kondo-singlet amplitude, $b^*$, for a 
pole \cite{Auerbach-prl86,Millis87}
of the conduction-electron self-energy
in energy space:
\begin{eqnarray}
\Sigma({\bf k},\omega)
=\frac{(b^*)^2}{\omega-\varepsilon_f^*} \quad .
\label{sigma-pole}
\end{eqnarray}
Here the self energy is specified via the Dyson
equation: 
$G_c({\bf k},\omega) =
\left [ \omega-\varepsilon_{\bf k} -
\Sigma({\bf k},\omega) \right ]^{-1}$.
Correspondingly, in the Kondo lattice model, 
the conduction-electron Green's function contains two poles, 
respectively at energies
\begin{eqnarray}
E^{\pm}_{{\bf k}} &=& \frac{1}{2}\,\left [\,\varepsilon_{\bf k}+\varepsilon_f^*
\pm \sqrt{(\varepsilon_{\bf k}-\varepsilon_f^*)^2+4(b^*)^2} \, \right ] 
\quad .
\label{gc-two-poles2}
\end{eqnarray} 
They 
describe
the heavy fermion bands.
The nonzero $b^*$ 
specifies a Kondo resonance and
leads to
 a large Fermi surface, where both the local moments and conduction electrons contribute.
The quasiparticle
weight 
is $Z_L \propto (b^*)^2$ 
(Fig.\,\ref{fig:qp-Z}). The damping rate has the Fermi liquid 
$(k_{\rm B}T)^2$ and $E^2$ form.}

\end{minipage}
}

\clearpage
\newpage
\fbox{
\begin{minipage}{0.98\textwidth}
{For
the Kondo destruction phenomena,
we highlight
three key 
characteristics.
}
First, 
{in the Kondo-destroyed phases},
the Kondo-singlet amplitude {$b^*$} vanishes in the ground state. {Consequently, the poles in the conduction-electron self-energy disappear, leading to a small Fermi surface for which only the conduction electrons contribute. }

Second, 
{for} the Kondo destruction quantum critical point, 
the vanishing of the Kondo-singlet amplitude in the ground state implies that the weight of any Landau quasiparticle goes to zero. This can also be explicitly seen from the finite-size spectrum of the many-body excitations as determined by the numerical-renormalization-group (NRG) approach: the spectrum can no longer be fit in terms of a combination of any quasiaprticles~\cite{Kandala2022}.

Third, the Kondo destruction
quantum critical point
is
interacting (as opposed to being Gaussian). 
Thus, $k_{\rm B}T$ is the only energy scale. Accordingly, singular responses such as the local spin and charge susceptibilities have a dynamical Planckian ($\hbar \omega / k_{\rm B}T$) scaling. This has been seen both in the Kondo destruction fixed point of the Bose-Fermi Kondo/Anderson model as determined by a dynamical-large-$N$ (where the index $N$ appears in the spin channel) approach \cite{Cai-charge20.1,Zhu.04} and in the SU(2) case \cite{Cai-charge20.1,Kandala2022}.

We
note on the extra fixed points that exist beyond the $\epsilon$-expansion. It turns out that, for small $\epsilon$, the SU(2) Bose-Fermi Kondo/Anderson model has more fixed points beyond those that are accessed by the $\epsilon$-expansion method. 
Panel (c) shows the RG flow diagram when $\epsilon$ is sufficiently small (or $s=1-\epsilon$ is sufficiently large) \cite{Cai-SU2_19.1}. As $\epsilon$ further increases, the fixed points are pair-wise annihilated, and this has recently been understood analytically based on a $1/S$ expansion (where $S$ is the spin size) \cite{Hu22.fix}.
\end{minipage}
}

\clearpage
\newpage


\fbox{
\begin{minipage}{0.98\textwidth}

\includegraphics*[width=
1.0\textwidth]{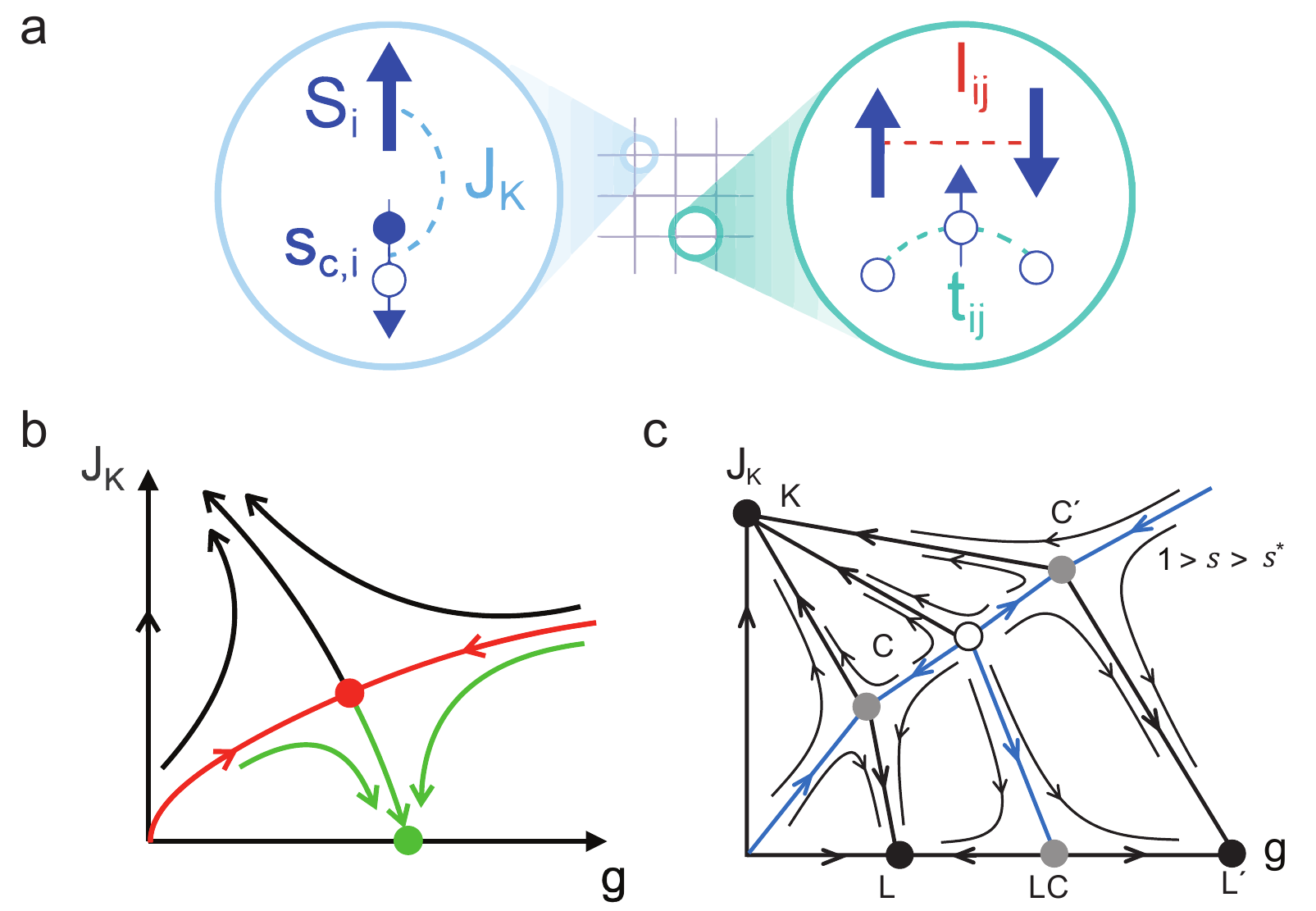}





{\bf Kondo lattice model and renormalization-group fixed points.}
{\bf a} Schematic illustration of the Kondo lattice model, which contains the Kondo coupling $J_K$ between local moments and conduction electrons, hopping parameters $t_{ij}$ between the conduction electrons, and RKKY interactions $I_{ij}$ between the local moments.
Adapted from Ref.\,\cite{Si-Nature}.
RG flow of the BFKM 
from studies based on {\bf b} an $\epsilon$-expansion \cite{ZhuSi} and {\bf c}
continuous time quantum Monte Carlo 
method \cite{Cai-SU2_19.1} and 
a large-$S$ expansion approach \cite{Hu22.fix} at $1>s>s^{*}$, where $s^{*}$ is a threshold value of the power-law exponent for the spectrum of the bosonic bath (see Eq.\,8).
The RG flow diagram illustrates two categories of stable fixed points: one associated with an infinite $J_K$, representing the Kondo phases, and the other characterized by $J_K=0$, indicating the local moment phases. 
An 
unstable fixed 
point
corresponds to 
a Kondo-destruction quantum critical point.
\end{minipage}
}


\end{document}